\documentclass[sn-mathphys-num]{sn-jnl}

\usepackage{graphicx}%
\usepackage{multirow}%
\usepackage{amsmath,amssymb,amsfonts}%
\usepackage [latin1]{inputenc}
\usepackage{mathrsfs}%
\usepackage[title]{appendix}%
\usepackage{xcolor}%
\usepackage{textcomp}%
\usepackage{manyfoot}%
\usepackage{booktabs}%
\usepackage{algorithm}%
\usepackage{algorithmicx}%
\usepackage{algpseudocode}%
\usepackage{listings}%

\begin{document}

\title[Article Title]{Effects of group size and noise on cooperation in population evolution of dynamic groups}


\author*[1]{\fnm{Hong-Bin} \sur{Zhang}}\email{zhanghb@suda.edu.cn}

\author[1]{\fnm{Deng-Ping} \sur{Tang}}\email{tangdengping@suda.edu.cn}

\affil*[1]{\orgdiv{School of Computer Science and Technology}, \orgname{Soochow University}, \orgaddress{\street{Shizhi Street}, \city{Suzhou}, \postcode{215006}, \state{Jiangsu}, \country{China}}}


\abstract{In a large population, the agents temporally form group of the Public Goods Game (PGG) one after another, and size of one group is randomly distributed at $g\in [g_l,g_h]$. Players in it have two strategies to be chosen to cooperate, or to defect for playing the PGG. Based on this structure we investigate the evolution of cooperation in PGG as a function of the noise level underlying strategy adoptions with the group size distribution. In the process of the evolution, the payoffs of cooperators(defectors) have related to the strategy selection implemented by the death-birth process. With a smaller group size $g$, different noise value $K$ induces different dynamic behavior with the increase of multiplication factor $r$. These cooperative behavior can be analytically solved. For a greater group size $g$, the system finally evolve to the bistable state(AllC and hybrid state with cooperators and defectors)starting from different initial concentrations of cooperation. The numerical computation also fit well with the simulation results.
We focus on the bistable phenomenon in the equilibrium for a given value of $K$ and a greater group size when the intrusion of one C(D)-player into D(C)-population is investigated from upwards(downwards) branch after experiencing a unstable cooperative evolution. Here, the hysteresis phenomenon can be observed. Further, when one D-player invades into the C-population, the critical value $r_c$ can be obtained in the simulation and the mathematical relationship between the critical value $r_c$ and noise $K$ is analysed.}

\keywords{Dynamic formed group, Public goods, Noise, Noise}



\maketitle

\section{Introduction}\label{sec1}

The rule of natural selection that is proposed by Darwin favors the selfish individuals who obtain the maximum benefit at the expense of the others\cite{Axelrod,Maynard,Hofbauer,Nowak}. For example, in a competing population, the individuals are divided into the cooperators and defectors, those cooperators are willing to contribute money so as to get benefits for the whole population while the defectors (also called free riders) don't pay anything but share the collections with the formers. Obviously, to be the defectors are rational and they will be winners since they get more benefit than the cooperators. Therefore, the defective behavior will be favored\cite{Hofbauer,Szabo1}. So the selection of cooperation in nature is not an obvious behavior. However, there is growing cooperation appeared in nature and society.

By now the evolutionary game theory is widely used to understand the competition behavior of agents in nature\cite{Nowak2,Galam17}. The Public Goods game (PGG) has been widely used to investigate the local interactions of multi-player systems\cite{Santos5}. In the group game, $g$ players forming one group, each in it can execute to cooperate, or to defect ($D$). A cooperator expends a cost $c$ so as to get a benefit $rc$ for the group,  where $r$ is the multiplication factor. Assume that $n$ cooperators are in the group, then a total gained benefit $nrc$ will be shared by all the players in the group. Therefore, the defectors have higher net payoffs since they do not pay extra costs. When $g$ players of formed groups enter into the PGG in a large population, for $1<r<g$ \cite{Hauert3}, the situation of the cooperators are worse off, and a defective population will evolve to a dominate position.

 The network effect that the competition environment of the population has been localized will lead to the emergence of cooperation among selfish individuals\cite{Maynard,Hamilton,Wilson,Trivers,Nowak3,Hauert,Hauert2}. Szolnoki proposed a model that individuals participate a PGG on regular graphs of different group size, through which the interaction structure of effective overlapping triangles can be structured among agents, and accordingly the cooperative effect on the noise $K$ dependence of the cooperative evolution has been discussed\cite{Szolnoki11}. Perc,Szolnoki,et al also spotted that the sphere of the mixture phase of C(D) players is enlarged with the growth of group size\cite{Perc,Szolnoki12,Galam15}. Obviously, it provides defectors different paths to invade the region of cooperators. In \cite{Broom}, Broom proposed that the increased variability of group size can favour cooperation evolution of PGG. Janssen gives a mechanism of dynamic-persistence of cooperation which players in a population forming one group can move toward other groups due to the influence of relative fitness of players in a population, resulting in the group evolution over time that further affect the cooperation in the population\cite{Janssen12}. Stivala, Zhang and et al suggested that the players have not perfect rationality, embedding the noise in the evolution of cooperation can affect the survival of cooperators\cite{Stivala,Zhang2,Chen2,Galam1}. Players selected from a well-mixed population randomly forming group engage in the PGG of constant group size, the defectors getting better payoffs on average\cite{Xu1}. Hauert and Kaiping et al.
  proposed that the update selection can be implemented by the death-birth processes in the population dynamics \cite{Hauert3,Kaiping1}. In this process, one cooperator (defector) to give the death is followed by the birth of a random agent, relating to the probability of reproduction of a C (D)-agent. The evolution of interaction of agents of temporally dynamic formed group will affect the average payoffs of C(D)-agents respectively in the population. The average payoffs also determine the the probability of reproduction of C(D)-agents. In finite population evolution of dynamic formed groups, the parameter $s$ is thought as the length of interactive time scale among agents and the strategy update is implemented by the death-birth process \cite{Roca}. In the situation, the parameter $s$ has an prominent impact on the probability of ALLC population in the evolutionary game\cite{Axelrod,Johnson1,Johnson2}.
 The size distributions of temporally formed group is also considered and play an important role on the formation of totally cooperative population when the strategy selection follows the imitation process \cite{Zheng,Xu}.

In what follows, we suggest an interesting comprehension of cooperative evolution when introducing the noise $K$ dependence on dynamic formed group at a fixed time slot $s=1$ in \cite{Xu1}. Different type of formed group with \cite{Szolnoki11,Perc,Szolnoki12,Galam15,Janssen12}, a coupling of the noise level and dynamic formed group size at random distribution gives an no-trivial dynamic phenomena and an new comprehension of cooperative evolution. We consider the total number of players in the population always remains unchanged before one turn begins. The probability of the offspring players inheriting the strategies of their parents. This reproduction probability is characterized by the Fermi function with a noise effect. The death-birth process is applied in our model. It is related to the total payoffs of cooperators(defectors). Compared with previous essays when the noise $K$ is introduced, in addition to the simulation results we can obtain, an analytical method related to the noise $K$ is also provided. The details are as follows.

\section{Model}\label{sec2}

We consider a large population of $M$ players attending the PGG game of dynamic formed group by one after the other until all players are dispatched to different groups in one turn. In the situation, interaction between unfixed agents of one dynamic formed group will happen. In the PGG, each type of agents which choose to cooperate, or to defect, will gain their their payoffs respectively. Here, in one dynamic formed group, assume that there are $g$ agents and $n$ cooperators attending the PGG, after one round the whole profits of cooperators in this group is given by

\begin{equation}
P_c (g,n) = (\frac{{rn}}{g} - 1)n,
\end{equation}

Meanwhile, after one round the whole profit of defectors in this group is given by:

\begin{equation}
P_d (g,n) = \frac{{rn}}{g} (g-n),
\end{equation}

\noindent Where $r$ ($>1$) is the multiplication factor. When one agent choose to cooperate, he must pay the extra cost which is taken as $c=1$. In this paper, the total payoffs that the two type agents accumulated after one round will depend on the strategy selection process, which is described with the death-birth process. In a time step, there are two phases: at the first one, the time scale of one strategy update is focused, which is same to that of one interaction of agents engaging in a PGG, and here $g$ players are selected from $M$ population to form one group following the random size distribution $g\in [g_l,g_h]$. The initial payoffs of cooperators and defectors are set as $P_c$=$P_d$=0. After the game, the profits that the two type of agents (cooperators and defectors) are still denoted with $P_c$ and $P_d$ respectively. The second stage is the death-birth process. The two-type agents compete to birth an offspring who inherits his father's strategy with a certain probability. The reproduction probability $w_c$ of the cooperators is given by $w_c  = 1 /[1+e^{(P_d  -  P_c )/K}]$ where $K$ is the noise value. The learning mechanism helps the offsprings inheriting a dominant strategy in the evolutionary process. The reproduction probability of the defectors is $w_d$=1-$w_c$. The process randomly make an agent die, and give birth to a new offspring with certain probabilities. The C(D) feature of a new offspring subjects to the reproduction probability $w_c$($w_d$). So far, after the death-birth process, one evolutionary time step is over and a dynamic formed group is dismissed, entering the next round.

In this paper, $M=10^5$ players are selected to proceed the Monte-Carlo simulation in which one Monte-Carlo step (MCS) consists of $10^5$ time steps. When the system is in equilibrium state, we get the average cooperative frequency $f_c$ by taking an average of $10^4$ MCS.

\section{Theory and Results}\label{sec3}
The frequency of cooperation $M_c/M$ is thought to weight up the yardstick of a system's performance, in which $M_c$ is the number of cooperators at time step $t$. The term of $h(g)$ expresses the size of dynamic formed group according to the random distribution, and let $\sum\nolimits_{g = g_l }^{g_h} {h(g)}  = 1$ with $g\in[g_l,g_h]$. In the $i$-th turn, we assume the probability that the there are $g_i$ agents and $n_i$ cooperators in one group is given by  \cite{Galam,Galam16}
\begin{equation}
p(g_i ,n_i ) = h(g_i )\left( {\begin{array}{*{20}c}
   {g_i }  \\
   {n_i }  \\
\end{array}} \right)f_c^{n_i } (1 - f_c )^{g_i  - n_i } , \label{eq:p}
\end{equation}
where $p(g_i ,n_i )$ in Eq.~(\ref{eq:p}) give an expression of the binomial distribution for the probability of having $n_i$ cooperative agents in a group with the size of $g_i$. After the $i$-th round game of PGG, the accumulated profits of C and D agents is given by:
\begin{equation}
P_{tot}( g_i ,n_i ) =P_{c}( g_i ,n_i )+P_{d}( g_i ,n_i )= (r-1)n_{i}.
\end{equation}
In general, the dynamic equation can be obtained:
\begin{equation}
\frac{{df_c }}{{dt}} = \frac{1}{M}(\sum\limits_{g = g_l }^{g_h } {\sum\limits_{n = 0}^g {p(g,n) w _c (g,n)} }  - f_c ). \label{eq:dfdt_s1}
\end{equation}
We will set the right side equal to zero, a static solution of $f_c$ can be obtained. If $g_h$ ($g_h<5$) has a relatively small value, analytical solution may be obtained. For example, when $g\in[1,3]$ and $h(g)$ follows an uniform distribution, the static solution of Eq.(5) can be obtained by solving the following univariate cubic equation:
\begin{equation}
af_c^3+bf_c^2+cf_c+\frac{1}{2}=0. \label{eq:dfdt_s1}
\end{equation}
where
\begin{equation}
a=-\frac{1}{6}+\frac{1}{3(1+e^{\frac{3-3r}{K}})}+\frac{1}{1+e^{\frac{6-2r}{3K}}}+\frac{1}{1+e^{\frac{3+r}{3K}}}. \label{eq:dfdt_s1}
\end{equation}
\begin{equation}
b=\frac{1}{3}(2-\frac{2}{1+e^{\frac{1}{K}}}+\frac{1}{1+e^{\frac{2-2r}{K}}}+\frac{3}{1+e^{\frac{6-2r}{3K}}}-\frac{6}{1+e^{\frac{3+r}{3K}}}). \label{eq:dfdt_s1}
\end{equation}
\begin{equation}
c=-2+\frac{2}{3(1+e^{\frac{1}{K}})}+\frac{1}{3(1+e^{\frac{1-r}{K}})}+\frac{1}{1+e^{\frac{3+r}{3K}}}. \label{eq:dfdt_s1}
\end{equation}

As we can see that the theoretical curves fit well to all the simulation data (see Figure 1)

\begin{figure}[h]
\centering
\begin{minipage}[b]{0.6\textwidth}
\includegraphics[width=\textwidth]{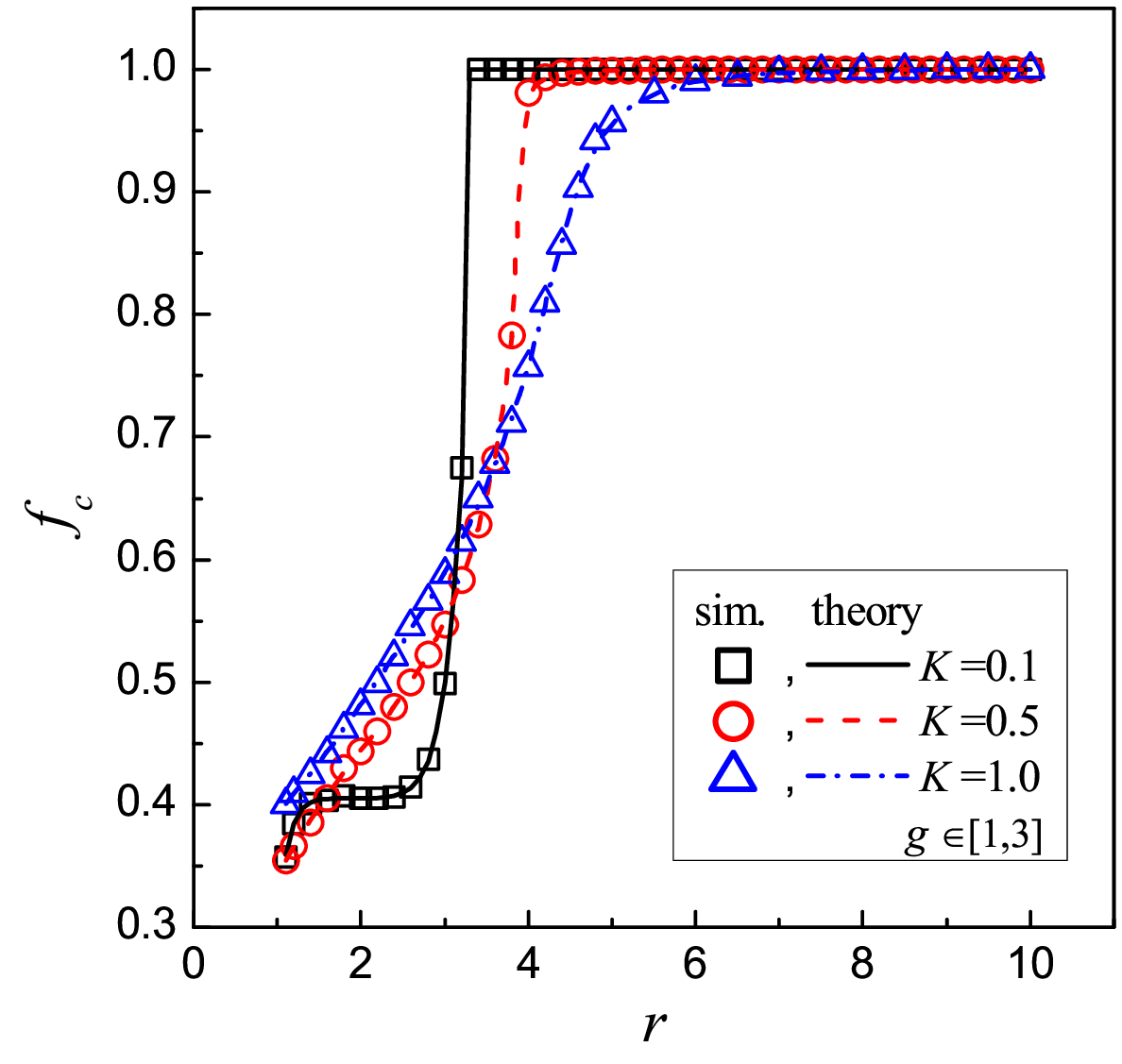}\\
\end{minipage}
\caption{(color online) The dependence of the frequency of cooperation $f_c$ on the multiplication factor $r$ at different noise strength $K$. The size of formed group $g\in[1,3]$ is set with the initial frequency of cooperation $f_{c,i}=0.5$. The marks are gained by simulations, but the curves are obtained by analytical solution.}  \label{fg:X1}
\end{figure}

The noise level $K$ will affect the frequency of cooperation. In Figure 1, it is noted that the frequency of cooperation $f_c$ has related to different values of $K$ when $g\in[1,3]$. For example, for the multiplication factor $r<3$, a greater noise value of $K$ promotes the frequency of cooperation, and the noise strengthen it in this region; however, when $r>3$, a smaller noise value of $K$ favours the cooperative behavior, and AllC state is easier to be achieved. Here, for $r>3$, the noise value of $K$ inhibits the increase of cooperative frequency. In Figure 1, we obtain that the theoretical solution fits well with the results of computer simulation. If the noise level $K$ is not controlled, Xu and Boom have indicated that greater group variability will affect cooperation frequency\cite{Xu1,Broom}. When the size distribution of formed group $g$ has been enlarged, e.g.,for $g_u>5$, it benefits the the survival of cooperators(see Figure 2). Here, the system exhibits solutions with multiple stable states. Due to the adoption of the probability of reproduction with Fermi form, from different initial concentrations of cooperation when $g\in[1,5]$, $r=3.5$ and $K$=0.1, the system takes $10^5$ time steps  to evolve to the mixed state of common existence of C agents and D agents, or to the AllC state. However, the phenomenon of  multiple stable states has not appeared in \cite{Xu1} where the probability of reproduction in it is $v_c=P_c/P_{tot}$ instead of that of Fermi form. Although there is no analytical solution in Figure 2, the numerical computation are still highly consistent with the simulation.

\begin{figure}[h]
\centering
\begin{minipage}[b]{0.6\textwidth}
\includegraphics[width=\textwidth]{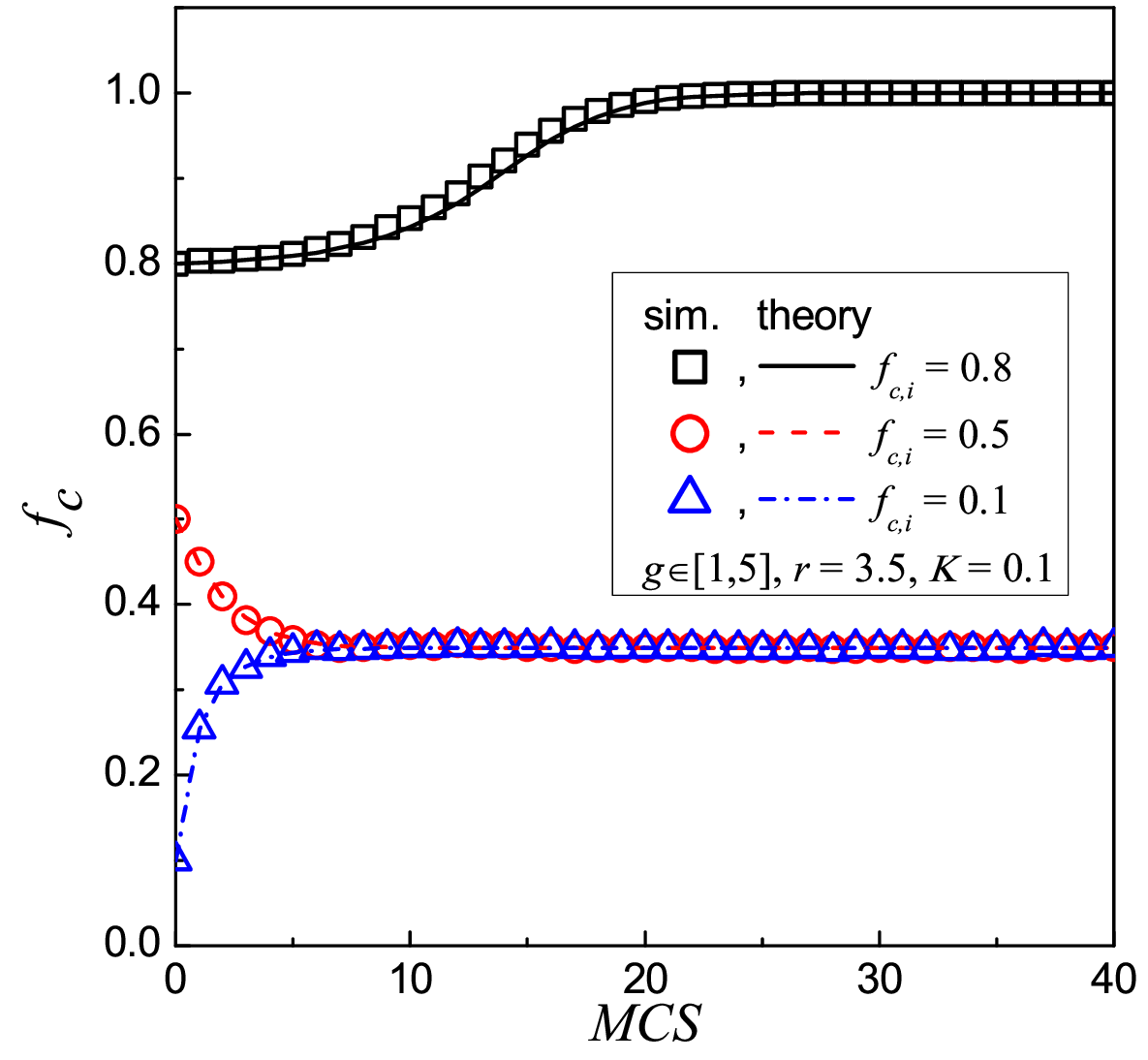}\\
\end{minipage}
\caption{(color online) After $20$ Monte-Carlo steps, the system converge to one stable mixed state of C and D, or AllC state when $g\in[1,5]$, $r=3.5$ and $K$=0.1. The marks are gained by simulations, but the curves are gained by numerical computation.}  \label{fg:X2}
\end{figure}

For a given value of $K=1.0$ and $g\in[1,5]$, it shows that hysteresis phenomenon occurs in Figure 3 when the frequency of cooperation changes with the multiplication factor $r$. Eventually, the simulation has evolved to two stable solutions after experiencing an unstable period of cooperation evolution. We explore an explanation that one defector (cooperator) gradually invade the territory of opposite side on two branches, which are depicted with solid lines. What we are depicting two branches, one is an upwards branch, the other is a downwards branch. Between two branches, there is an unstable solution by a dashed line. In Figure 3, the upwards branch reveals that the frequency of cooperation goes up with the increase of $r$, as the other goes down with the decrease of $r$. In the dynamic process evolving to the upwards, or to the downwards branch, what we set the current initial cooperation frequency relating to $f_c$ in the equilibrium state with the value of $r$ of the previous process. In the upwards branch, e.g., for the minimum value of $r$=1.1, the initial value of $f_c$ is close to 0.00001. In the another branch, for a bigger value of $r$=9.0, the initial value of $f_c$ is set to 0.99999.

\begin{figure}[h]
\centering
\begin{minipage}[b]{0.6\textwidth}
\includegraphics[width=\textwidth]{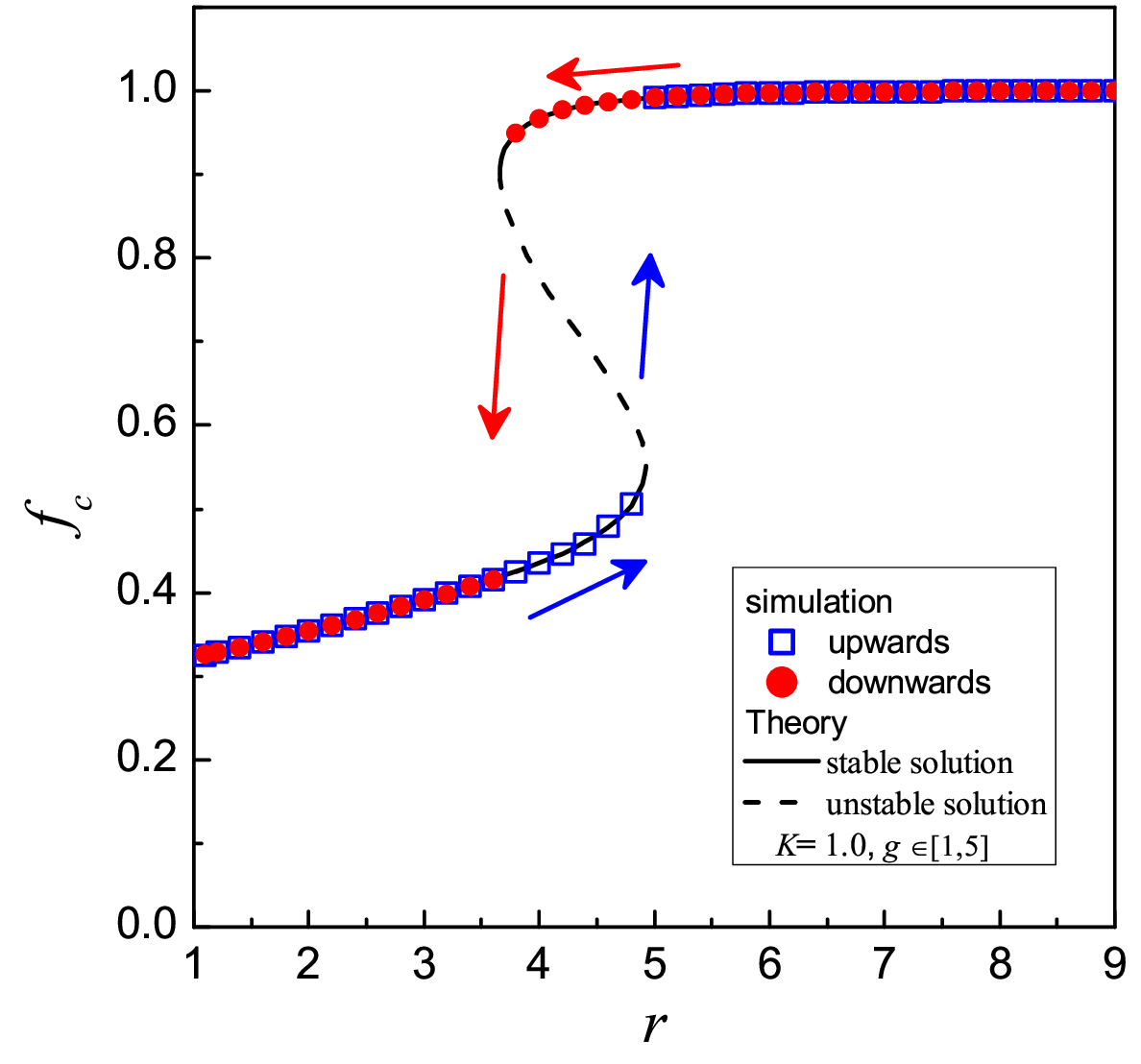}\\
\end{minipage}
\caption{(color online) The dependence of the frequency of cooperation with bistable state on the multiplication factor starting from different initial frequency of cooperation. We take a random size distribution with $g\in[1,5]$ and $K=1.0$.}  \label{fg:X3}
\end{figure}

Starting from different initial cooperation frequencies, the system will evolve to ALLC state or mixture state of $C$ and $D$. Thus, an interesting question here is whether multi-stable solution of the system may arise if we consider that one C-player invades the territory of AllD population, or one D-player invades the sphere of AllC population at different multiplication factor $r$. In Figure 4, for $K=0.1$ and $g\in[1,5]$, we investigate how one C-agent invades the area of all defectors by the description of square dots moving upward (upwards branch), and how one D-agent bursts into the territory of all cooperators from another side by those triangulation dots moving downward (downwards branch). In this situation, there must be an intrusion point from upwards branch or downwards branch. The theoretical solution is depicted by a curve where the dash line stands for non-stationary solution.

\begin{figure}[h]
\centering
\begin{minipage}[b]{0.6\textwidth}
\includegraphics[width=\textwidth]{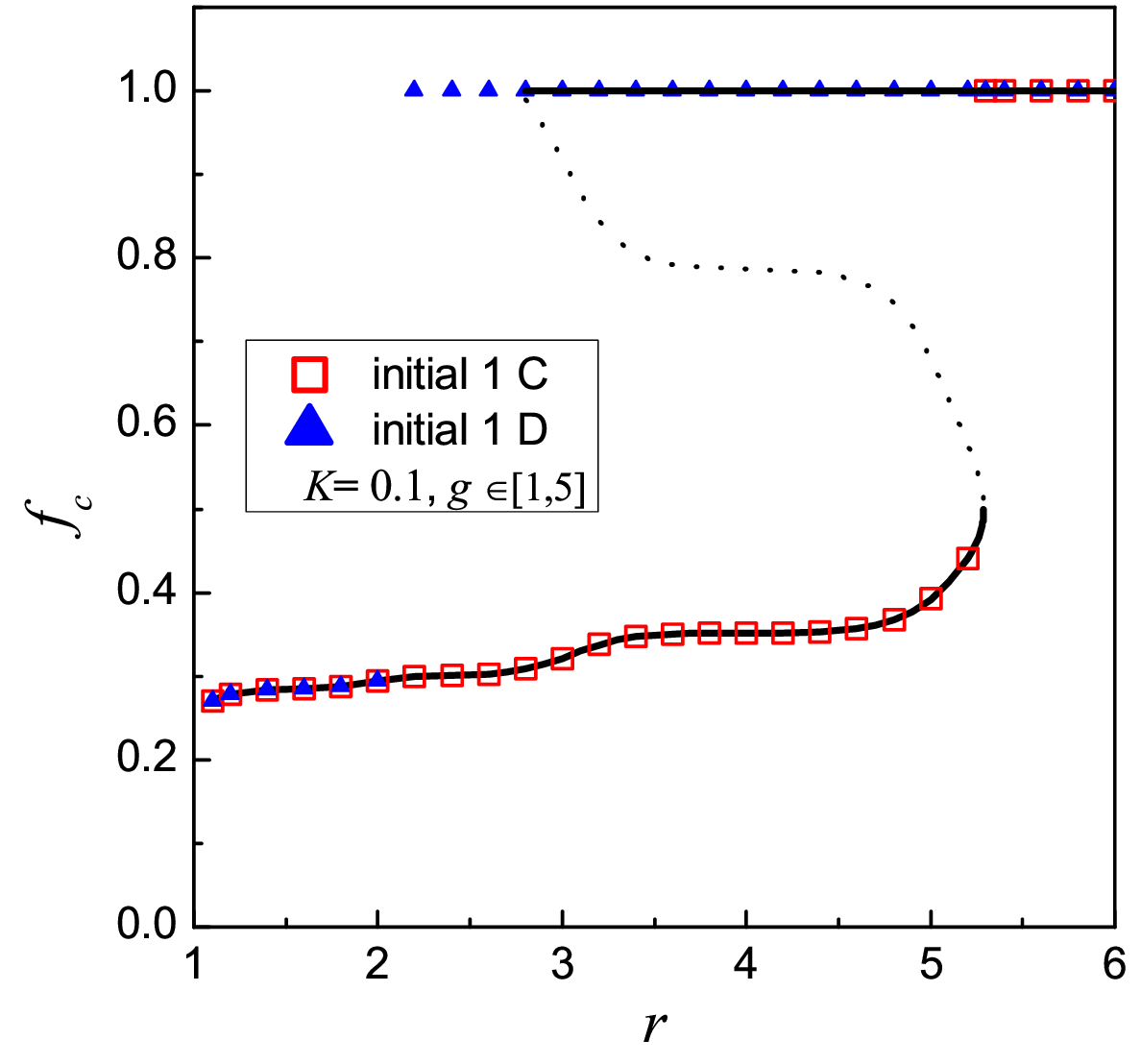}\\
\end{minipage}
\caption{(color online) When $K=0.1$ and $g\in[1,5]$, we can obtain results similar to Figure 3. The symbols describe the process that (upwards(downwards)) branches are shaped by one C(D)-agent invading the territory of Defectors(Cooperators).}  \label{fg:X4}
\end{figure}

From Figure 4, it can be seen that the intrusion to ALLC population by one D-player has related to the multiplication factor $r$. Further, to investigate the critical value for one D-player invading ALLC population, we averaged 1000 initial state process when the noise value of $K=0.1$ is given. More detailed explantation we make can be seen in the Figure 5. It gives a phase diagram of a single D-agent invading the region of AllC (see triangular point of Figure 4). We divide the phase diagram into three regions, e.g., region (I),region (II) and region(III). For a lower value of $r$, an AllC population can be certainly invaded by one D-agent in the region (I). In the region (II), one D-agent has chances to invade C-agents population, and the same is true for one C-agent. However, the region(III) is that AllC population cannot be invaded by one D-agent. In the region(III), we suppose that the system only has one defective agent and $M$-1 cooperative agents. If one cooperator to be born has higher probability than one defector, then the last defector will disappear. At this time, $0$ or $1$ defector is to be generated in the population related to the PGG of formed group and $g_l>1$. The probability of one defector has occurred in the group with size $g$ is given by $p_{1d}=g/M$. Obviously, the probability that zero defector lives in the system is $p_{0d}=1-g/M$. After experiencing an interaction, the probability $\beta_c$ that one cooperator to be born is given by
\begin{equation}
\begin{aligned}
\beta _c & = \sum\limits_{g_j  = 1 }^{g_h }\frac{1}{g_h} *{\left(\left(1-\frac{g_j}{M}\right)*{ {\frac{1}{1+e^{\frac
{0-\left(\frac{r*g_j}{g_j}-1\right)*g_j}{K}}}+\frac{g_j}{M}*\frac{1}{1+e^{\frac{\frac{r*(g_j-1)}{g_j}*1-\left(\frac{r*(g_j-1)}{g_j}-1\right)*(g_j-1)}{K}}}} } \right)}\\
&-\frac{M-1}{M},\label{eq:betac}
\end{aligned}
\end{equation}

Eq.~(5) provide the reference method to deduce the above equation.  Accordingly, the the probability $\beta_d$ for one defector to be born is given by
\begin{equation}
\begin{aligned}
\beta _d& = \sum\limits_{g_j  = 1 }^{g_h }\frac{1}{g_h} *{\left(\left(1-\frac{g_j}{M}\right)*{ {\frac{1}{1+e^{\frac
{\left(\frac{r*g_j}{g_j}-1\right)*g_j-0}{K}}}+\frac{g_j}{M}*\frac{1}{1+e^{\frac{\left(\frac{r*(g_j-1)}{g_j}-1\right)*(g_j-1)-\frac{r*(g_j-1)}{g_j}*1}{K}}}} } \right)}\\
&-\frac{1}{M},\label{eq:betac}
 \end{aligned}
\end{equation}

Compared Eq.~(10) with Eq.~(11), we get $\beta_c+\beta_d$=0. Obviously when $\beta_c>0$, $\beta_c>\beta_d$ and the AllC state will happen. For $g\in[1,5]$, a more simpler form is given as follows when $r>1$ and $K=0.1$,
\begin{equation}
\beta _c=-\beta _d =\frac{1}{5}(Q1+Q2+Q3+Q4+Q5)-\frac{1}{M} \label{eq:beta}
\end{equation}
where
\begin{equation}
Q_1=\frac{1}{2M}
\label{eq:beta}
\end{equation}

\begin{equation}
Q_2=\frac{2}{M}\frac{1}{1+e^{\frac{-1}{K}}}
\label{eq:beta}
\end{equation}

\begin{equation}
Q_3=\frac{3}{M}\frac{1}{1+e^{\frac{(\frac{2}{3}*r)-2}{K}}}
\label{eq:beta}
\end{equation}

\begin{equation}
Q_4=\frac{4}{M}\frac{1}{1+e^{\frac{(\frac{3}{2}*r)-3}{K}}}
\label{eq:beta}
\end{equation}

\begin{equation}
Q_5=\frac{5}{M}\frac{1}{1+e^{\frac{(\frac{12}{5}*r)-4}{K}}}
\label{eq:beta}
\end{equation}

Based on Eqs. (12)-(17), the critical value of $r_{invade}$ for invading the AllC population can be obtained by setting $\beta_c = 0$. In the simulations, we take $M=10^5$ and $K=0.1$, then the theoretical calculation of $r_{invade}$ is given by 2.75, which is close to that obtained in the simulations.

\begin{figure}[h]
\centering
\begin{minipage}[b]{0.6\textwidth}
\includegraphics[width=\textwidth]{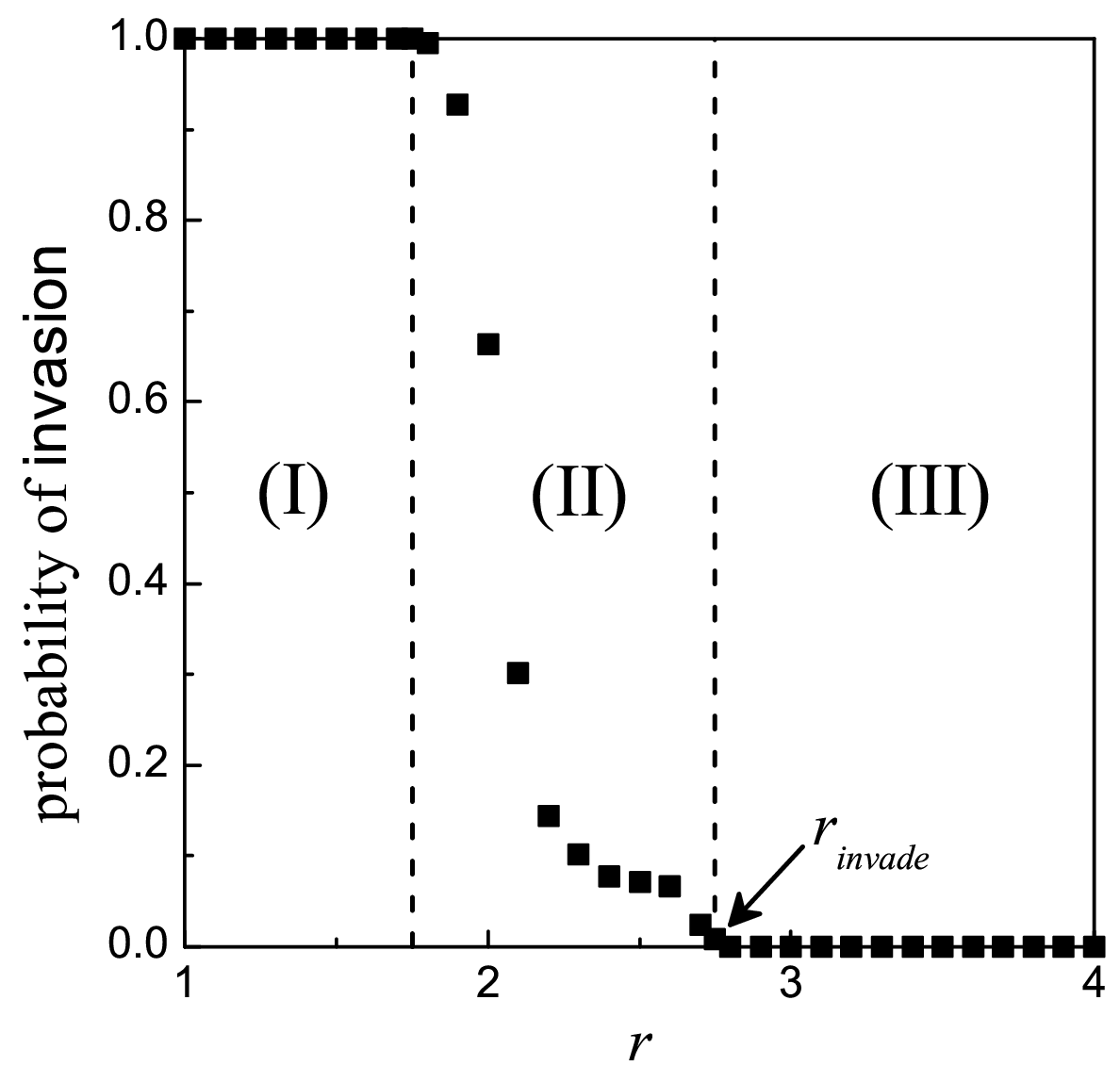}\\
\end{minipage}
\caption{(color online) Phase diagram of a single D-agent invading the region of AllC. The phase indicates three states: high probability of invasion, moderate probability of invasion, and impossible invasion with the description of
region (I),region (II) and region(III) respectively. All points can be obtained when $M=10^5$, $K=0.1$ and $g\in[1,5]$ by going through average 107 MCS steps.}  \label{fg:X5}
\end{figure}

\section{Summary}\label{sec4}

Summarizing, in a large population, we investigate the evolution of cooperation in the public goods game on dynamic randomly formed group with the group size $g$ and the noise level $K$. The payoffs of cooperators and defectors depends on the strategy update in the death-birth process. For the multiplication factor $r<3$, a greater value of the noise $K$ advances the cooperation frequency; but for $r>3$, the cooperation frequency is inhibited by the noise value $K$. For a smaller group size $g\in[1,3]$, different noise strength $K$ will induce different dynamic behavior of cooperation, and we also obtain analysis solutions fitted well with the simulation results. We also discuss that the situation with a greater group size, e.g.,for the noise value $K$=0.1, $g\in[1,5]$ and $r=3.5$, starting with different initial concentrations of cooperation, AllC state and mixed state of common existence of C and D agents can be observed. We find that the simulation is highly consistent with the results of numerical computation. Similarly, the bistable state in the system also exists when $K=1.0$. Further, for $K=1.0$ we study the evolution of cooperation dependence on different multiplication factor $r$, it can be observed that there exist a sphere of unstable solutions between the two stable extremes. The phenomenon of hysteresis effect can be used to give a comprehensible explanation for one C(D)-agent to invade the territory of the D(C) population with the change of the multiplication factor $r$. For $K$=1.0, on the downwards branch, one by one D-agent invade the region of C-population with decrease of the multiplication factor $r$, and another phase of mixed state of existence of C and D agents is eventually reached at the larger size of $g\in[1,5]$ which has verified in\cite{Szolnoki12}; however on the upwards branch, similarly one by one C-agent invade the sphere of D-population with the increase of $r$, the system finally approaches to the AllC state. For $K=0.1$ and $g\in[1,5]$, when a small value of $r$ is set, one defective agent easily invades the sphere of cooperators when a larger mean size of $g\in[1,5]$ yielding a larger sphere of phase favors the spread of defective strategy(see Figure 2)\cite{Szolnoki12}; but when above the critical value of $r_{invade}$, e.g, for $r_{invade}=2.75$, he is hard to invade the territory of cooperators. Also, the mathematical relationship between critical value of $r_c$ and noise $K$ is given and discussed when one D-player invade to AllC population.




\section*{Author contribution statement}

Hong-Bin Zhang: Simulation and calculation, Writing and original draft. Deng-Ping Tang: Methodology, Writing, review,
editing.

\section*{Declaration of competing interest}

The authors declare that they have no known competing financial interests or personal relationships that could have
appeared to influence the work reported in this paper.

\section*{Data Availability Statement}

No data was used for the research described in the article.



\end{document}